\documentclass[10 pt, journal,table,xcdraw, hyphens, pdftex, cmex10,  colorlinks,breaklinks]{IEEEtran}

\makeatletter
\def\ps@headings{%
	\def\@oddhead{\mbox{}\scriptsize\rightmark \hfil \thepage}%
	
	\def\@evenhead{\scriptsize\thepage \hfil \leftmark\mbox{}}%
	
	\def\@oddfoot{}%
	
	\def\@evenfoot{}}
	
\makeatother
\usepackage{cite}
\usepackage[colorlinks]{hyperref}
\usepackage{soul}
\usepackage{longtable}
\usepackage{float}  
\usepackage{epsfig,bm,amsmath,amssymb,graphicx,setspace}
\usepackage[numbers,sort&compress]{natbib}
\usepackage{color,placeins}
\usepackage{multirow}
\usepackage{algorithm}
\usepackage{algpseudocode}
\usepackage{colortbl}
\usepackage{color}
\usepackage{dblfloatfix}
\usepackage{turnstile}
\usepackage{multicol}
\usepackage{array}

\usepackage{enumerate}
\usepackage{turnstile}
\usepackage{subcaption}
\usepackage{arydshln,paralist}
\usepackage{soul,tabularx,booktabs}
\usepackage{csquotes}
\usepackage{multirow}
\usepackage{adjustbox}
\usepackage{float}

\usepackage{caption}
\usepackage{subcaption}

\usepackage{xcolor}
\usepackage[colorlinks]{hyperref}
\usepackage{verbatim}
\usepackage[colorinlistoftodos]{todonotes} 
\usepackage{rotating}
\definecolor{usethiscolorhere}{rgb}{0.86666,0.78431,0.78431}
\usepackage{tikz}
\usetikzlibrary{mindmap}
\usetikzlibrary{calc,positioning}
\usepackage{lipsum,adjustbox}
\usetikzlibrary{decorations.pathmorphing}

\usepackage{pifont}

\makeatother

\usepackage{amsmath}

\definecolor{lime}{HTML}{A6CE39}

\usepackage[pdftex]{graphicx}

\DeclareGraphicsExtensions{.pdf,.jpeg,.png,.eps}

\usepackage{algorithm,algpseudocode}
\usepackage{caption}
\usepackage{epstopdf}
\usepackage{cite}
\usepackage{caption}
\usepackage{subcaption}

\usepackage{amsfonts}
\usepackage[none]{hyphenat}
\usepackage{url}

\usepackage{hyperref}
\usepackage{comment}
\usepackage[euler]{textgreek}
\usepackage{epsfig,bm,amssymb,graphicx,setspace,amsfonts}
\usepackage{atbegshi}
\AtBeginDocument{\AtBeginShipoutNext{\AtBeginShipoutDiscard}}
\usepackage{algorithm}
\usepackage[ruled,vlined,algo2e]{algorithm2e}
\usepackage{color}
\usepackage[T1]{fontenc}
\usepackage{textgreek}
\usepackage{amsthm}
\usepackage{array}
\usepackage{balance}
\usepackage{enumerate}

\begin{document}

\title{Beyond Reality: The Pivotal Role of Generative AI in the Metaverse}

\author{ Vinay Chamola, \textit{Senior Member, IEEE},  Gaurang Bansal, Tridib Kumar Das, Vikas Hassija, Naga Siva Sai Reddy, \\ Jiacheng Wang, Sherali Zeadally, \textit{Senior Member, IEEE},   Amir Hussain, \textit{Senior Member, IEEE},  \\
Richard Yu, \textit{Fellow, IEEE} , Mohsen Guizani, \textit{Fellow, IET} and Dusit Niyato, \textit{Fellow, IEEE}}

\thanks{Vinay Chamola and Siva Sai are with the Department of Electrical and Electronics Engineering, BITS-Pilani, Pilani Campus, 333031, India. (e-mail: vinay.chamola@pilani.bits-pilani.ac.in, siva.sai@pilani.bits-pilani.ac.in).}
\thanks{Gaurang Bansal with the Department of Electrical and Computer Engineering, National University of Singapore, Singapore. (e-mail:gaurang@u.nus.edu).}
\thanks{Vikas Hassija and Tridib Kumar Das are with School of Computer Engineering, Kalinga Institute of Industrial Technology, KIIT, Bhubaneswar, India (email: vikas.hassijafcs@kiit.ac.in, 2128106@kiit.ac.in)}
\thanks{Sherali Zeadally is with College of Communication and Information, University of Kentucky, Lexington, KY (email: szeadally@uky.edu)}
\thanks{Amir Hussain is with the School of Computing, Edinburgh Napier University, Scotland, UK (email: A.Hussain@napier.ac.uk)}
\thanks{Jiacheng Wang and Dusit Niyato are with School of Computer Science and Engineering, Nanyang Technological University, Singapore (e-mail: jiacheng.wang@ntu.edu.sg, dniyato@ntu.edu.sg).}
\thanks{Mohsen Guizani is with the Department of Computer Science, Qatar University, Qatar (e-mail: mguizani@ieee.org).}

\maketitle

\begin{abstract}
Imagine stepping into a virtual world that's as rich, dynamic, and interactive as our physical one. This is the promise of the Metaverse, and it's being brought to life by the transformative power of Generative Artificial Intelligence (AI). This paper offers a comprehensive exploration of how generative AI technologies are shaping the Metaverse, transforming it into a dynamic, immersive, and interactive virtual world. We delve into the applications of text generation models like ChatGPT and GPT-3, which are enhancing conversational interfaces with AI-generated characters. We explore the role of image generation models such as DALL·E and MidJourney in creating visually stunning and diverse content. We also examine the potential of 3D model generation technologies like Point-E and Lumirithmic in creating realistic virtual objects that enrich the Metaverse experience. But the journey doesn't stop there. We also address the challenges and ethical considerations of implementing these technologies in the Metaverse, offering insights into the balance between user control and AI automation. This paper is not just a study, but a guide to the future of the Metaverse, offering readers a roadmap to harnessing the power of generative AI in creating immersive virtual worlds.
\end{abstract}
  

\section{INTRODUCTION}
\IEEEPARstart{M}{etaverse} is a concept that describes a virtual universe, an interconnected network of immersive digital environments where users can interact with each other and digital content \cite{metaverse}. It merges physical reality with digital virtuality, offering a seamless blend of real-world experiences and virtual interactions. The Metaverse has garnered significant hype due to its potential to revolutionize various aspects of human life, including entertainment, social interactions, commerce, education, and more \cite{metaverse}. It promises to provide limitless opportunities for creativity, exploration, and collaboration.

The usefulness of the Metaverse lies in its ability to create immersive and interactive experiences that transcend the limitations of physical reality. It enables people to connect, communicate, and engage with others and digital content in ways that were previously unimaginable. The Metaverse opens up new avenues for entertainment, business, commerce, and intercommunication by allowing users to participate in virtual games, attend virtual events, explore virtual worlds, and more \cite{metaverse}. It also facilitates social interactions, enabling people to meet, socialize, and collaborate in virtual spaces, allowing for new business models and revenue streams \cite{metaverse}.

The future of the Metaverse is undeniably intertwined with the advancements and innovations in artificial intelligence (AI) \cite{metaverse}. AI is poised to play a pivotal role in shaping and defining the future trajectory of the Metaverse. AI is indispensable for the Metaverse due to its ability to enhance user experiences, automate content creation, and enable intelligent interactions. Among AI algorithms, generative AI is particularly pivotal in generating realistic virtual content, personalizing experiences, and creating dynamic virtual worlds \cite{generativeAI}.

Generative AI refers to a branch of artificial intelligence that focuses on creating new and original content, such as images, text, music, or virtual environments \cite{generativeAI}. Thus, generative AI plays a crucial role in the evolution of the Metaverse, enabling novel experiences and enhancing user engagement. For example, generative AI algorithms can procedurally generate virtual worlds within the Metaverse. By leveraging sophisticated algorithms, these systems can create vast and diverse landscapes, structures, and environments \cite{proceduralAI}. Each virtual world becomes a unique and immersive experience for users, as generative AI ensures dynamic and ever-evolving content. Another use case is the generation of personalized avatars and characters in the Metaverse. AI algorithms can generate lifelike and diverse virtual beings, complete with customizable appearances, traits, and behaviors \cite{avatarGeneration}. The integration of generative AI within the Metaverse opens up a range of exciting possibilities.

\begin{figure*}
\centering
\begin{subfigure}{.5\textwidth}
  \centering
  \includegraphics[width=\linewidth]{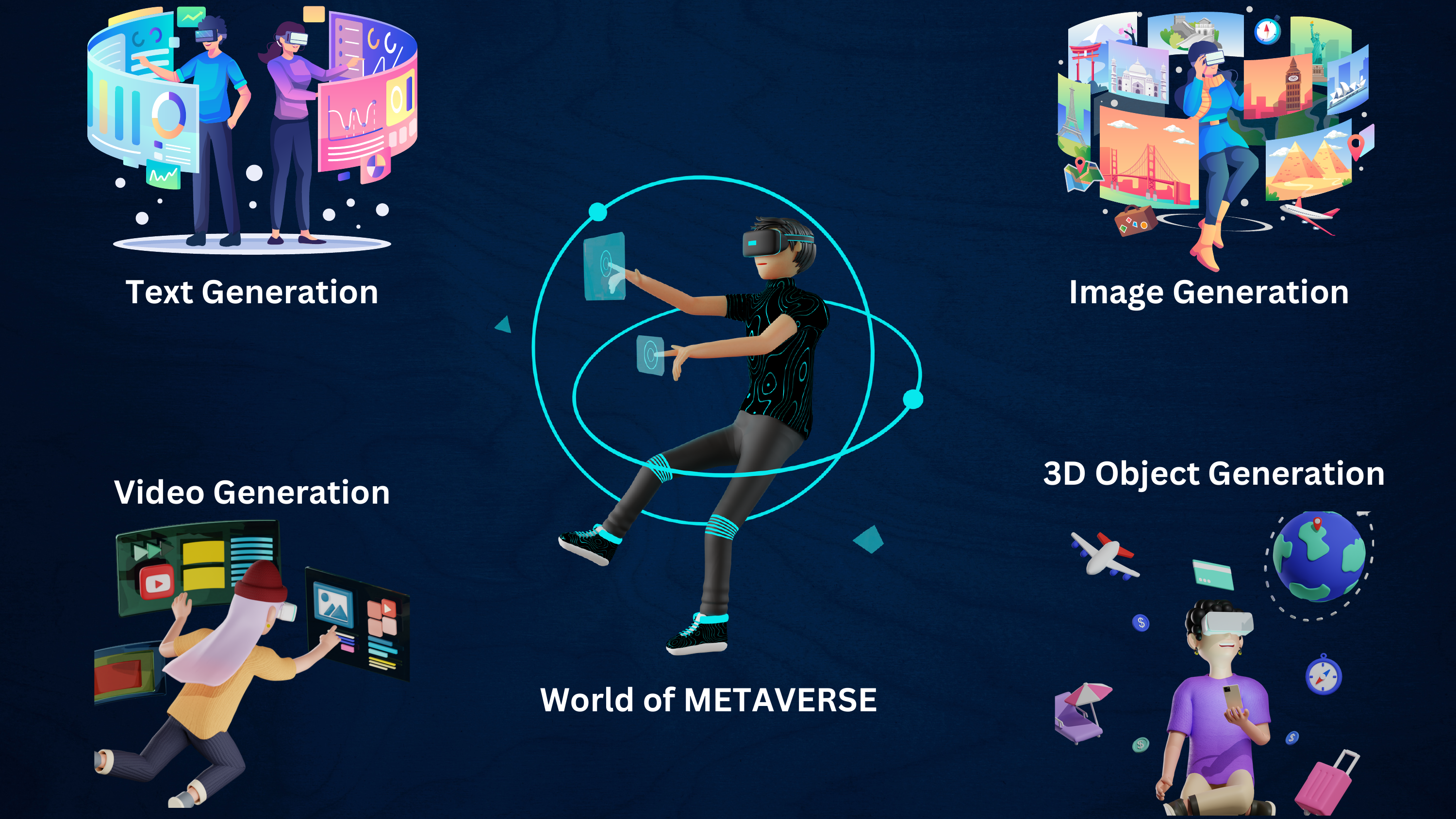}
  \caption{Domains in world of metaverse}
  \label{fig:sub1}
\end{subfigure}%
\begin{subfigure}{.5\textwidth}
  \centering
  \includegraphics[width=\linewidth]{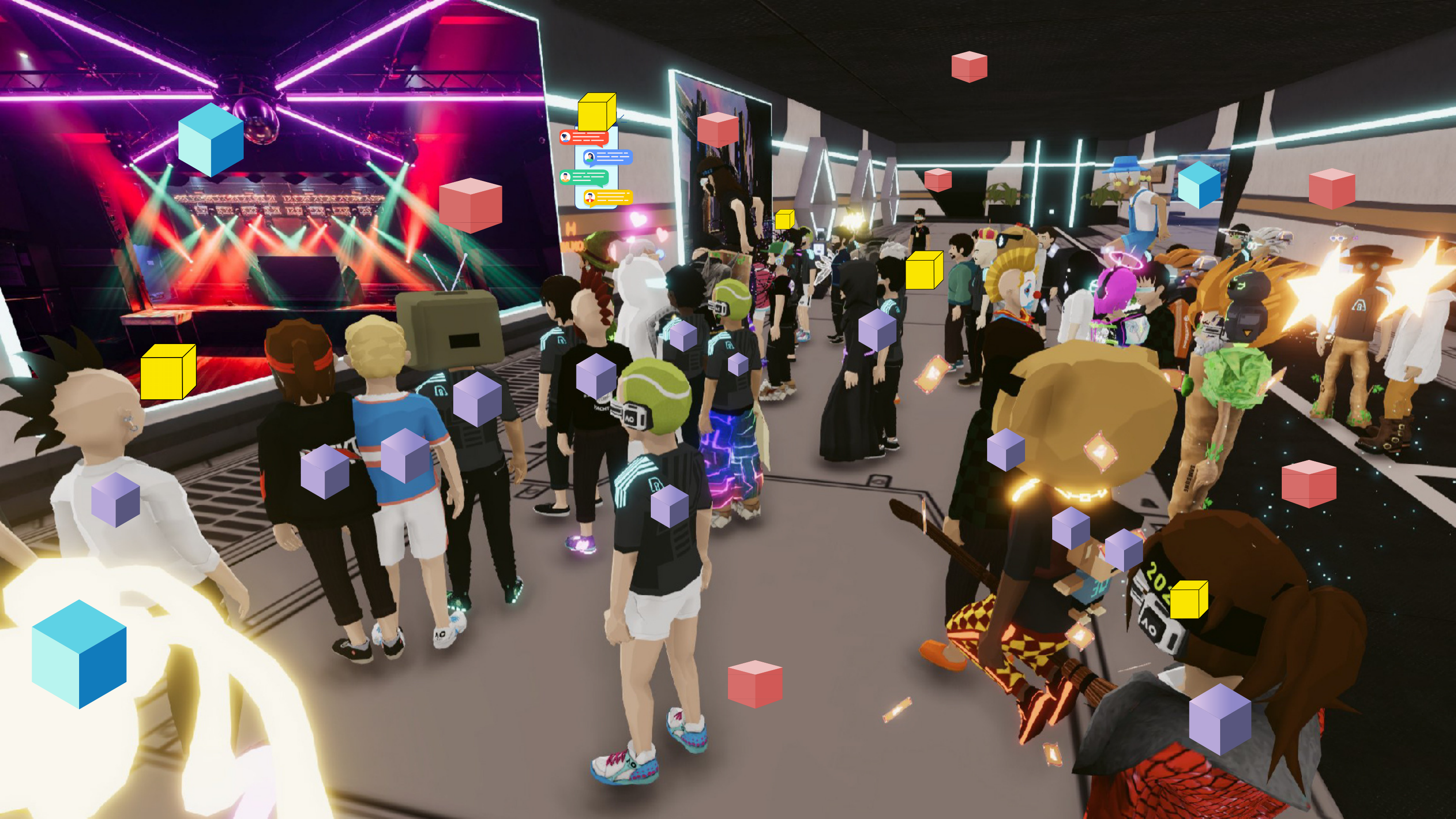}
  \caption{Divisions highlighted by cubes in Metaverse.}
  \label{fig:sub2}
\end{subfigure}
\caption{Domains in World of Metaverse. (a) describes different domains, while (b) highlights different domains in metaverse marked with different color cubes. Yellow cube denotes text, blue for video, red for image and purple for 3D objects.}
\label{fig:test}
\end{figure*}

In conclusion, the Metaverse represents a transformative concept that combines virtual and physical reality, offering unprecedented opportunities for human interaction and engagement \cite{metaverse}. AI, especially generative AI, plays a crucial role in shaping the Metaverse by enabling immersive experiences, content generation, and personalized interactions \cite{generativeAI}. The integration of AI in the Metaverse holds great promise for the future of entertainment, commerce, social interactions, and beyond \cite{metaverse, generativeAI, proceduralAI, avatarGeneration}. 


The generation of metaverse can be classified into four major domains, so becomes easier to understand and analyze the specific areas where generative AI technologies can be applied to enhance and enrich the overall experience within the virtual realm as shown in Figure 1. We highlight how metaverse consists of the different domains marked with different color cubes. Yellow cube denotes text, blue for video, red for image and purple for 3D objects.

\begin{enumerate}

\item \textbf{Text Generation:} This domain focuses on the generation of textual content within the Metaverse. It includes applications such as chatbot conversations, virtual world descriptions, interactive narratives, and text-based interactions between users.

\item \textbf{Image Generation:} Image generation involves the creation of visual content within the Metaverse. It encompasses tasks like generating virtual landscapes, objects, avatars, textures, and other graphical elements that contribute to the immersive visual experience.

\item \textbf{Video Generation:} Video generation pertains to the creation of dynamic visual content within the Metaverse. It involves generating animated scenes, virtual tours, interactive videos, and other forms of moving imagery that enhance the visual storytelling and engagement.

\item \textbf{Object Generation:} Object generation focuses on the creation of virtual 3D objects and assets within the Metaverse. It encompasses the generation of virtual furniture, architecture, vehicles, characters, and other interactive elements that populate the virtual environment.

\end{enumerate}

Similarly, we classify the generative models into 4 major model categories as:

\begin{enumerate}
\item \textbf{Variational Autoencoders (VAEs):} VAEs are generative models that learn a compressed representation (latent space) of the input data. They consist of an encoder network that maps the input data to the latent space and a decoder network that reconstructs the input data from the latent representation. VAEs are commonly used for tasks like image generation and data synthesis.

\item \textbf{Generative Adversarial Networks (GANs):} GANs consist of a generator network and a discriminator network that play a minimax game. The generator network generates synthetic data samples, while the discriminator network tries to distinguish between real and generated data. GANs are widely used for generating realistic images, videos, and other data types.

\item \textbf{Transformers:} Transformers are attention-based models that excel in capturing long-range dependencies in sequential data. They have been successfully applied to various generative tasks, including natural language processing, language translation, and image generation. Transformers leverage self-attention mechanisms to process input sequences and generate high-quality outputs.

\item \textbf{Autoregressive Models:} Autoregressive models generate data sequentially, where each element in the sequence is conditioned on the previous elements. Autoregressive models are widely used for generating text, images, and other sequential data types.
\end{enumerate}

This paper explores the impact of generative AI in the evolution of the Metaverse by examining the intricate interactions between four types of generative AI models and domains of the Metaverse. By delving into the capabilities and applications of different Models, alongside the domains of text generation, image generation, video generation, and object generation, we aim to provide an in-depth understanding of how generative AI fuels the development and transformation of the Metaverse. Through comprehensive analysis and evaluation, this paper sheds light on the potential of generative AI to revolutionize the Metaverse, paving the way for immersive, interactive, and personalized virtual experiences.

\begin{figure}
    \centering
    \includegraphics[width = \columnwidth]{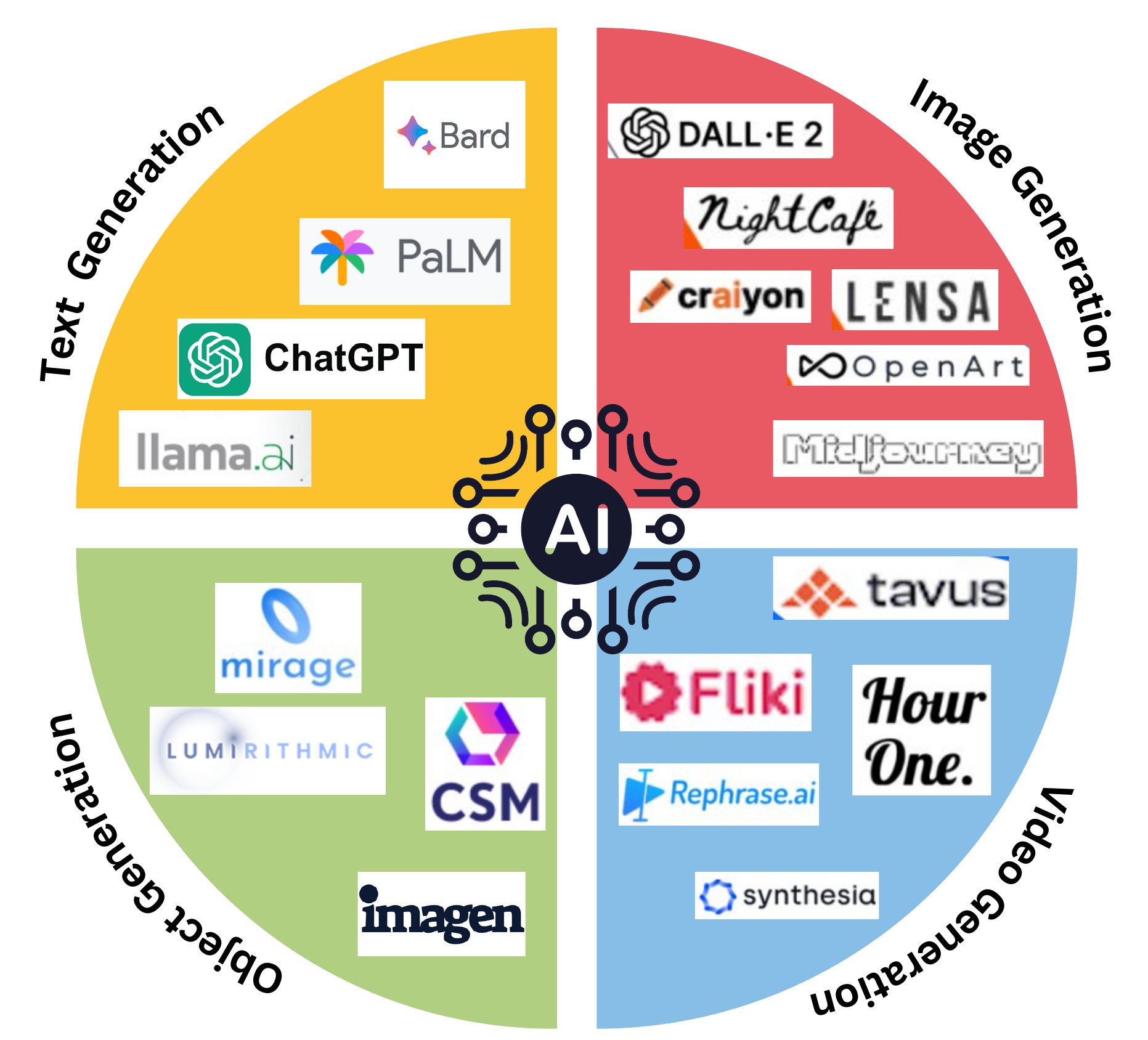}
    \caption{Different Generative AI technologies impacting domains of Metaverse.}
    \label{fig:enter-label}
\end{figure}

\begin{table*}
\scriptsize
\centering
\caption{Generative AI Technologies in Metaverse Domains}
\label{tab:generative-ai-metaverse}
\resizebox{\textwidth}{!}{
\begin{tabular}{|>{\columncolor[HTML]{C0C0C0}}p{0.08\textwidth}|p{0.08\textwidth}|p{0.1\textwidth}|p{0.4\textwidth}|p{0.2\textwidth}|}
\hline
\rowcolor[HTML]{C0C0C0}
\textbf{Metaverse Domains} & \textbf{Technology} & \textbf{Generative AI Model} & \textbf{Technical Details} & \textbf{Application} \\ \hline
\textbf{Text Generation} & XLNet & Transformer &  Leverages the Transformer architecture to generate context-aware text based on input sequences. & Natural language generation, chatbot responses, interactive storytelling \\ \cline{2-5}
& PaLM & Transformer & Captures dependencies between linguistic units at different granularities to capture contextual relationships in text. & Text completion, language generation, document summarization.\\ \cline{2-5}
& GPT & Transformer & Transformer-based language model that generates text based on input prompts. & Natural language generation, Language translation, Text summarization, Writing assistance \\ \cline{2-5}
& BARD & Autoregressive Model & Generates text by estimating the distribution of the next word given the previous words. & Content creation, Creative writing, Dialog systems, Text generation \\ \cline{2-5}
& Llama & Variational Autoencoders & Employs hierarchical latent spaces to capture structured representations of text to generate coherent and contextually relevant text based on specific conditions or prompts. & Text generation, content creation, dialog systems \\ \hline
\textbf{Image Generation} & DALL-E & Variational Autoencoders & Uses a discrete codebook to generate images by sampling and decoding from the learned latent space. DALL-E can produce novel and creative images based on given prompts or conditions. & Art generation, image synthesis, content creation \\ \cline{2-5}
\multirow{6}{*}{} & CRAIYON & GANs & Combines style transfer and image synthesis techniques to generate visually appealing and novel images. It allows for control over style and content during the generation process. & Art generation, style transfer, content creation \\ \cline{2-5}
& NightCafe & GANs & Generates high-quality landscape and scenery images based on user input or predefined themes. It focuses on producing realistic and immersive natural scenes. & Landscape generation, scene synthesis, virtual environment creation \\ \cline{2-5}
& LENSA & Variational Autoencoders & Generates artistic and abstract images by learning latent space representations from diverse visual datasets. It allows for exploration and manipulation of the latent space to produce unique visual outputs. & Art generation, abstract image synthesis, creative exploration \\ \cline{2-5}
& Open Art & Variational Autoencoders & Employs variational autoencoders to generate diverse and creative artwork. It allows for the exploration of latent space to discover novel and unique artistic outputs. & Art generation, creative exploration, visual expression \\ \cline{2-5}
& MidJourney & GANs & MidJourney generates intermediate steps between given images, enabling smooth transition and interpolation in image synthesis. & Image morphing, visual effects, creative design. \\ \cline{2-5}
 & Stable Diffusion & GANs & Stable Diffusion  utilizes diffusion-based techniques to generate high-quality images by iteratively adding noise to an initial image. & Image synthesis, photo manipulation, creative design \\  \hline
\textbf{Video Generation} & Flicki & GANs &Generate dynamic and diverse video content based on user-defined styles and themes. & Video generation, creative storytelling, visual content creation \\ \cline{2-5}
& Runway & GANs & Provides an interface for managing data, training models, and generating video content. & Video synthesis, creative experimentation, artistic expression \\ \cline{2-5}
& Hour One & Variational Autoencoder & Enables the generation of realistic virtual actors for various applications, such as films, games, and virtual reality experiences. & Virtual actor creation, character animation, storytelling \\ \cline{2-5}
& Tavus & GANs & Generate visually appealing and novel video content based on given input and user-defined parameters. & Video synthesis, visual effects, creative content generation \\ \cline{2-5}
& Rephrase.ai & GANs & Rephrase.ai specializes in generating synthetic video content with realistic and customizable avatars. It allows users to animate virtual characters using text or voice inputs, enabling the creation of personalized video messages, tutorials, and more. & Avatar animation, virtual spokesperson, personalized video content \\ \cline{2-5}
& Synthesia & Variational Autoencoder & Create personalized videos by synthesizing realistic avatars that can speak and act like real people. It enables the automation of video production processes and the creation of dynamic video content at scale. & Personalized video messaging, marketing videos, automation \\ \hline
\textbf{3D Object Generation} & 3D-GAN & GANs & 3D-GAN captures the 3D geometry and appearance of objects by learning a latent space representation, allowing for the generation of diverse and detailed 3D objects. & 3D modeling, virtual object creation, and architectural design. \\ \cline{2-5}
\multirow{8}{*}{}& CSM & GANs & Enhances object detection by incorporating contextual information and past observations to improve accuracy and consistency. & Enhanced 3D object detection, context-aware perception \\ \cline{2-5}
& Mirage & Autoregressive Model & Employs techniques such as data augmentation, adversarial training, or synthetic data generation to improve the performance and robustness of object detection models. & Improved 3D object detection, data augmentation, model robustness  \\ \cline{2-5}
& ControlNet & Autoregressive Model & Separates the shape and pose variations in 3D object generation, enabling precise control over the generated objects' appearance and configuration. & Artistic design, 3D modeling, and virtual object creation. \\ \cline{2-5}
 & Imagen & GANs & Creates 3D objects from a single 2D image by leveraging a conditional generative network. It employs a conditional mapping from 2D images to 3D shapes, allowing for the reconstruction and generation of 3D objects. & 3D object reconstruction, content creation, and virtual reality assets. \\ \cline{2-5}
 & Point-E & GANs & Generates 3D point clouds by learning a conditional generative network and capturing the geometric structure of objects by mapping from a low-dimensional noise vector to a set of 3D points, enabling the synthesis of complex and detailed 3D object shapes. & 3D object synthesis, point cloud generation, and shape modeling. \\ \cline{2-5}
 & Lumirithmic & GANs & Generates realistic 3D objects with controllable lighting and material properties by incorporating lighting and material attributes into the 3D object generation process. & virtual object creation, lighting design, and computer graphics. \\ \cline{2-5}
 & ShapeNet & Variational Autoencoders & Models the shape and appearance of 3D objects using a VAE architecture, allowing for the generation of realistic and diverse 3D object samples. & 3D modeling, virtual object creation, and architectural design. \\ \cline{2-5}
 & DeepSDF & Variational Autoencoders  & Learns the underlying 3D geometry of objects by encoding their shape as an SDF, enabling the generation of high-quality 3D objects with precise geometric details. & 3D modeling, shape completion, and shape representation learning. \\ \hline
 \end{tabular}}
\end{table*}

\section{Transformative Power of Generative AI in the Metaverse}
Having discussed how the Metaverse comprises different domains, we uncover the potential applications and technical details of generative AI models, providing insights that can guide developers, researchers, and industry professionals in leveraging these technologies to enhance user experiences and drive innovation within the Metaverse \cite{generativeAIapplications}. In Table I, we provide an overview of generative AI technologies and their applications in various Metaverse domains (shown in Figure 2) from \cite{generativeAIapplications}. It covers various domains such as Text Generation, Image Generation, Video Generation, and 3D Object Generation, showcasing different models and their applications (shown in Figure 2).

In Text Generation \cite{gpt}, models like XLNet, PaLM, GPT, BARD, and Llama are featured. They utilize Transformer architectures, autoregressive models, and variational autoencoders for tasks such as natural language generation, chatbot responses, text completion, and interactive storytelling. Image Generation \cite{lensa} includes models like DALL-E, CRAIYON, NightCafe, LENSA, Open Art, MidJourney, and Stable Diffusion. They employ variational autoencoders and GANs to generate creative and realistic images for art generation, style transfer, landscape generation, and more. Video Generation \cite{flicki} introduces models such as Flicki, Runway, Hour One, Tavus, Rephrase.ai, and Synthesia. They leverage GANs and variational autoencoders for dynamic video content generation, virtual actor creation, and personalized video messaging. For 3D Object Generation \cite{mirage}, models like 3D-GAN, CSM, Mirage, ControlNet, Imagen, Point-E, Lumirithmic, ShapeNet, and DeepSDF are presented. These models utilize GANs, variational autoencoders, and other techniques for tasks like 3D modeling, shape completion, and enhanced object detection. The table offers a comprehensive overview of these technologies, enabling a better understanding of their capabilities and potential applications in the metaverse.

\section{Automated Text Generation in Metaverse}

We now explore the technical use cases of generative AI-based text generation in the metaverse, showcasing how it enhances different aspects of virtual experiences. From chatbot interactions and interactive storytelling to content generation and language translation, generative AI models pave the way for dynamic, personalized, and immersive textual interactions within virtual environments. 

\begin{itemize}
    \item \textbf{Chatbot Interactions}: Generative AI models, such as GPT (Generative Pre-trained Transformer), can power chatbots within the metaverse, enabling dynamic conversations and contextually relevant responses. For example, a virtual assistant chatbot in a virtual reality game can guide players through the gameplay, answer their questions, and engage in natural language-based interactions.
    
    \item \textbf{Interactive Storytelling}: Employing narrative generation techniques, generative AI models facilitate interactive storytelling in the metaverse. They generate coherent and engaging narratives based on user inputs and choices. For instance, in a virtual role-playing game, the story evolves based on the player's decisions, and the generative AI model generates text to reflect the consequences of those choices.
    
    
    \item \textbf{Language Translation}: Sequence-to-sequence models, like Google's Neural Machine Translation (GNMT), enable real-time language translation in the metaverse, allowing seamless communication across different languages. For instance, in a virtual conference with participants from different countries, generative AI can provide on-the-fly translation of text-based interactions, ensuring effective communication among attendees.
    
    \item \textbf{Personalized Virtual Assistants}: Generative AI models serve as personalized virtual assistants in the metaverse, adapting their responses to individual user preferences and providing tailored guidance. For example, a virtual assistant can learn user preferences, generate recommendations for virtual experiences, and provide personalized assistance throughout the metaverse journey.
    
    \item \textbf{Natural Language Interaction}: Models like OpenAI's ChatGPT enhance natural language interaction in the metaverse, enabling realistic and dynamic conversations with AI-driven avatars. For instance, in a virtual reality social platform, users can engage in lifelike conversations with virtual characters, allowing for a more immersive and interactive social experience.
    
    \item \textbf{In-Game Dialogues and Quests}: Generative AI-based text generation allows for dynamic in-game dialogues and quests in virtual worlds, with NPCs generating adaptive dialogues and quests based on user choices and progress. For example, non-playable characters in an open-world game can respond to the player's actions and generate quests that are tailored to their character's progression, providing personalized and engaging gaming experiences.
    
\end{itemize}

\begin{figure*}
    \centering
    \includegraphics[width = .9\textwidth]{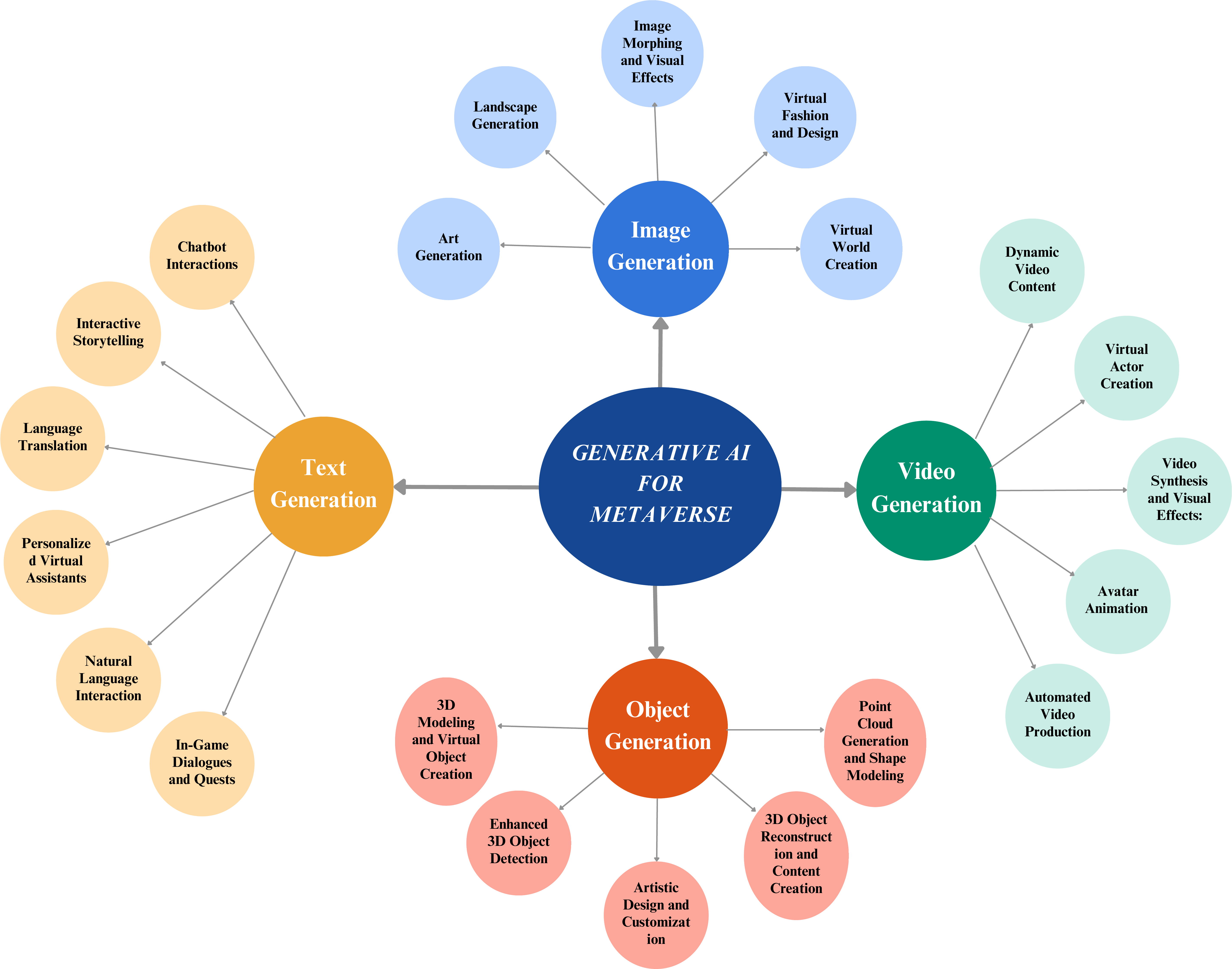}
    \caption{Applications of Generative AI in Metaverse.}
    \label{fig:enter-label}
\end{figure*}

\section{Image Generation and the Metaverse }

Generative AI-based image generation holds immense potential for enhancing the metaverse, a virtual realm of interconnected 3D environments. By leveraging advanced machine learning techniques, such as GANs (Generative Adversarial Networks) and variational autoencoders, generative AI models offer promising avenues for creating and enriching visual content within the virtual environment. Let us delve into the ways in which they can be effectively employed:

\begin{itemize}
\item \textbf{Art Generation}: Generative AI models like DALL-E and CRAIYON utilize variational autoencoders and GANs to generate unique and creative artwork in the metaverse. For example, artists in the metaverse can use generative AI to create virtual art galleries filled with AI-generated masterpieces, showcasing diverse artistic styles and themes.

\item \textbf{Landscape Generation}: Models such as NightCafe utilize GANs to generate high-quality landscape and scenery images in the metaverse. This enables the creation of realistic and immersive natural scenes, ranging from breathtaking landscapes to imaginative worlds, providing visually captivating environments for virtual experiences.

\item \textbf{Image Morphing and Visual Effects}: Models such as MidJourney utilize GANs to generate intermediate steps between given images, allowing for smooth transitions and seamless visual effects in the metaverse. For instance, in virtual reality applications, generative AI can create realistic image morphing and stunning visual effects during scene changes, enhancing the immersion and visual impact of the virtual experience.

\item \textbf{Virtual Fashion and Design}: Generative AI models can assist in virtual fashion and design within the metaverse. By generating clothing and fashion items, these models enable virtual designers to create and showcase virtual fashion lines, accessories, and personalized virtual wardrobes. This supports the growth of virtual fashion industries and enables immersive virtual commerce experiences.

\item \textbf{Virtual World Creation}: Generative AI-based image generation plays a crucial role in creating immersive virtual worlds in the metaverse. By generating realistic and diverse images of landscapes, buildings, and objects, these models contribute to the visual richness and believability of virtual environments. For example, generative AI can create virtual cities, forests, or futuristic settings, providing users with visually captivating and engaging virtual worlds to explore and interact with.

\end{itemize}


\begin{table*}[h]
\scriptsize
\centering
\caption{Open Issues and Future Work in Generative AI for the Metaverse}
\label{tab:open-issues-future-work}
\resizebox{\textwidth}{!}{
\begin{tabular}{|>{\columncolor[HTML]{C0C0C0}}p{2cm}|p{3cm}|p{6cm}|p{3cm}|p{2cm}|}
\hline
\rowcolor[HTML]{C0C0C0}
\textbf{Category} & \textbf{Open Issue} & \textbf{Solution Approach} & \textbf{Future Work} & \textbf{Application} \\
\hline
\textbf{Data} & Obtaining high-quality training data in the metaverse is challenging. & Develop advanced data collection techniques, data augmentation methods, and synthetic datasets that mimic the metaverse environment. & Improve data availability and quality for generative AI models in the metaverse. & Text, Image, Video, 3D Object Generation \\
\hline
\textbf{Realism }& Generating realistic and high-fidelity content is crucial for immersive experiences. & Advance model architectures, training algorithms, and unsupervised learning techniques to improve the realism and fidelity of generated content. & Enhance the visual quality and realism of generative AI models in the metaverse. & Image, Video Generation \\
\hline
\textbf{Content Control} & Ensuring proper content control and moderation in the metaverse is essential. & Develop robust content filtering and moderation systems, leveraging automated algorithms and human oversight to prevent the dissemination of harmful or offensive generated content. & Ensure responsible and safe content generation in the metaverse. & Text, Image, Video Generation \\
\hline
\textbf{Ethics} & Generative AI in the metaverse raises ethical concerns regarding privacy, consent, and ownership. & Establish clear ethical guidelines, regulations, and legal frameworks to address these concerns and ensure responsible use of generative AI technologies. & Promote ethical and responsible practices in the use of generative AI models in the metaverse. & Text, Image, Video, 3D Object Generation \\
\hline
\textbf{Computational Efficiency} & Generative AI models can be computationally demanding. & Develop more efficient algorithms, model compression techniques, and hardware optimizations to reduce computational requirements and enable real-time generation in the metaverse. & Enhance the efficiency and speed of generative AI models for real-time applications in the metaverse. & Image, Video, 3D Object Generation \\
\hline
\textbf{Interoperability} & Ensuring interoperability and standardization across different metaverse platforms is crucial. & Develop open standards, APIs, and frameworks that facilitate interoperability and encourage the sharing and transfer of generative AI models and content. & Promote seamless integration and collaboration among generative AI models in the metaverse. & Text, Image, Video, 3D Object Generation \\
\hline
\end{tabular}}
\end{table*}

\section{Video Generation in Metaverse}

Generative AI-based video generation revolutionizes the metaverse, transforming it into a realm of captivating visual experiences. Harnessing the power of advanced machine learning, these cutting-edge models unlock the potential for dynamic and diverse video content creation within virtual environments as shown in following: 

\begin{itemize}
\item \textbf{Dynamic Video Content}: Models like Flicki and Runway generate diverse video content based on user-defined styles, supporting creative storytelling and immersive virtual experiences. For instance, in a virtual conference, generative AI can dynamically generate video content with different visual styles for presentations, enhancing engagement and visual appeal.

\item \textbf{Virtual Actor Creation}: Models such as Hour One and Rephrase.ai synthesize realistic virtual actors for films, games, and virtual reality experiences. These actors, with customizable appearances and expressions, enhance character animation and interactive virtual environments. For example, in a virtual training simulation, generative AI can create virtual actors that deliver lifelike scenarios to trainees.

\item \textbf{Video Synthesis and Visual Effects}: Generative AI models like Tavus and Stable Diffusion generate visually appealing video content with customizable visual elements and effects. They enhance visual storytelling and content creation within the metaverse. For instance, in a virtual art gallery, generative AI can synthesize video content that dynamically displays artwork with various visual effects and transitions.

\item \textbf{Avatar Animation and Personalized Video Content}: Models like Rephrase.ai and Synthesia generate personalized video content with realistic avatars. Users can animate virtual characters using text or voice inputs, creating personalized video messages or tutorials. For example, in a virtual customer support system, generative AI can produce video responses with personalized avatars to address customer queries.

\item \textbf{Automated Video Production}: Generative AI models, such as Synthesia, automate video production processes within the metaverse. They synthesize realistic avatars that can speak and act like real people, enabling the efficient generation of personalized video messaging and marketing videos. For instance, in a virtual marketing campaign, generative AI can automate the creation of personalized video advertisements with virtual spokespersons.
\end{itemize}


\section{Looking at the Prospects of Generative 3D }

3D object generation holds significant potential for enhancing the metaverse, a virtual universe where people interact and engage in various activities. By employing advanced algorithms and machine learning techniques, generative AI models can create realistic and diverse 3D objects that populate virtual environments. These models enable:

\begin{itemize}
\item \textbf{3D Modeling and Virtual Object Creation}: Generative AI models facilitate 3D modeling and virtual object creation within the metaverse, supporting architectural design, virtual object libraries, and immersive virtual experiences. For example, in a virtual reality game, users can create and customize their virtual objects, such as buildings, vehicles, and furniture, using generative AI-based 3D modeling tools.

\item \textbf{Enhanced 3D Object Detection}: Models enhance object detection by incorporating contextual information and past observations, enabling enhanced 3D object detection and context-aware perception within the metaverse. For instance, in an augmented reality application, generative AI can improve the detection and tracking of virtual objects in real-world environments, enhancing the user's augmented reality experience.

\item \textbf{Artistic Design and Customization}: Models enable precise control over the appearance and configuration of generated objects, supporting artistic design, 3D modeling, and virtual object creation within the metaverse. For example, artists can use generative AI-based tools to create and customize virtual sculptures, allowing for intricate details and personalized artistic expression.

\item \textbf{3D Object Reconstruction and Content Creation}: Generative AI models can create 3D objects from 2D images, enabling the reconstruction and generation of 3D objects from visual references, supporting content creation, virtual reality assets, and 3D object synthesis in the metaverse. For instance, a virtual museum can use generative AI to transform 2D images of artifacts into interactive 3D models, providing an immersive and educational experience.

\item \textbf{Point Cloud Generation and Shape Modeling}: Models generate 3D point clouds by capturing the geometric structure of objects, facilitating the synthesis of complex and detailed 3D object shapes, and contributing to point cloud generation, 3D object synthesis, and shape modeling within the metaverse. For example, in a virtual design environment, generative AI can generate detailed point clouds of architectural structures, allowing architects and designers to explore and refine virtual spaces.
\end{itemize}

\section{Case Study}
\begin{figure*}
    \centering
    \includegraphics[width = .9\textwidth]{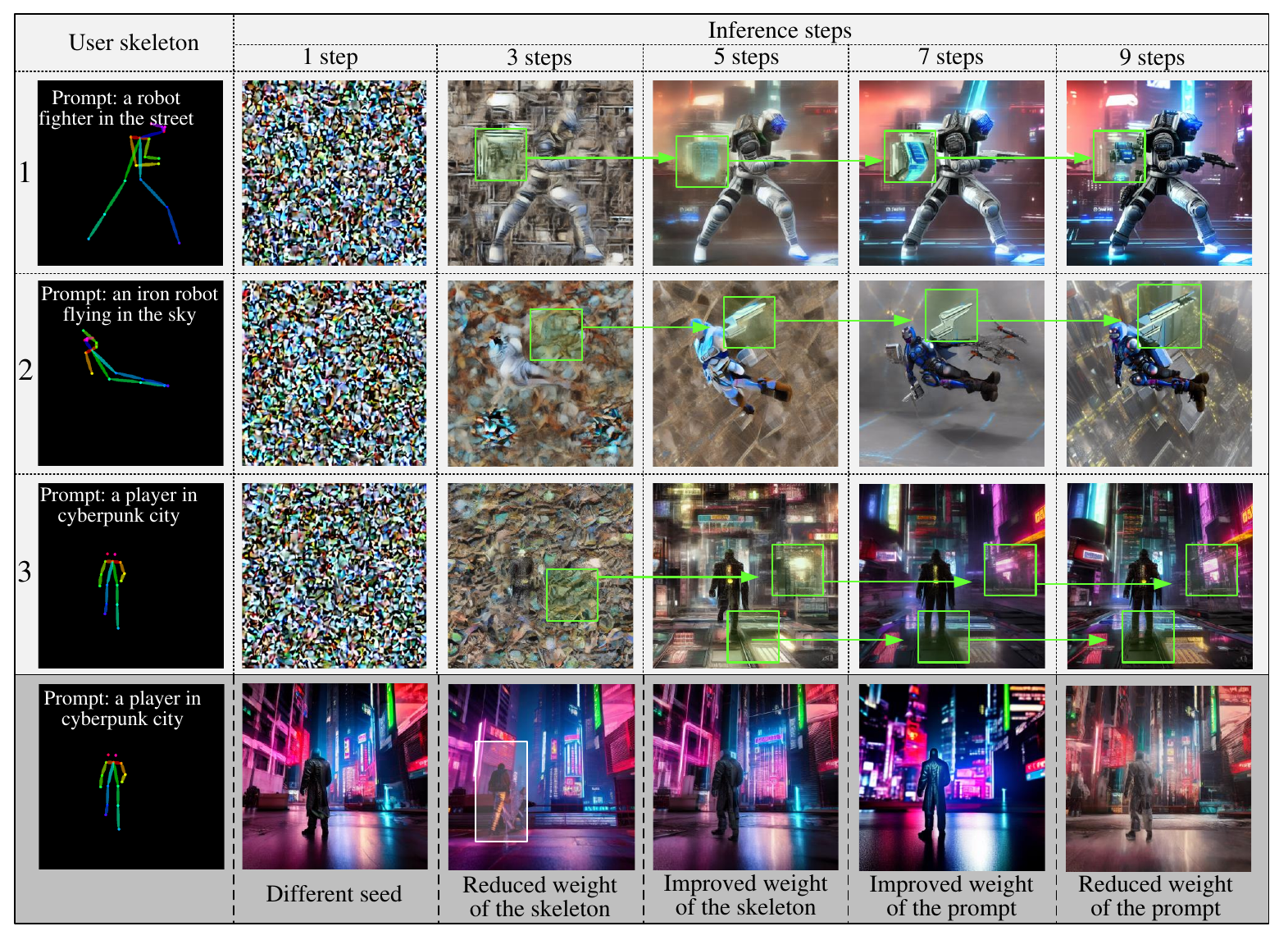}
    \caption{The results of utilizing a diffusion model for user avatar generation in the Metaverse.}
    \label{fig:case2}
\end{figure*}

In previous sections, we have explained the role of generative AI in the Metaverse from various perspectives. In this section, we demonstrate how generative AI specifically supports the Metaverse using the example of virtual avatar generation for Metaverse players. Concretely, we utilize a camera to capture the video stream about the user in the physical world and convert it into skeleton using OpenPose~\cite{cao2017realtime}. After that, the extracted skeleton is combined with user's prompts to generate virtual avatar in the Metaverse for users using a diffusion model~\cite{zhang2023adding}, which is one of the generative AI models\footnote{The code is available at https://github.com/lllyasviel/ControlNet}. The experimental results are presented in Fig.~\ref{fig:case2}.

From the results, we can see that the diffusion model can effectively generate avatars based on user's skeleton and prompts, and the quality of the generated avatars gradually improves with an increasing number of inference steps of the diffusion model. Specifically, when only one inference step is performed, the generated content is merely noise, as shown by the result corresponding to step1 in the Fig.~\ref{fig:case2}. As the number of inference steps increases, the avatar starts to gradually emerge and its posture aligns with the skeleton of the user. For example, in the results shown in the third row, as the number of steps increases, the limbs of the avatar are gradually fleshed out and the details of the background are also progressively refined, as indicated by the areas marked by the green boxes.

In addition, we analyze the impact of random seeds and the weights assigned to the skeleton and prompts on avatar generation. As shown by the figures in the fourth row in the Fig.~\ref{fig:case2}, given the same skeleton and prompt, the avatar and corresponding background generated vary when different random seeds are used. However, the posture of the avatar remains consistent with the provided skeleton. Moreover, when we adjust the weights of the skeleton and prompt in the avatar generation process, we can also obtain different results, as illustrated by the last four images in the fourth row. For instance, when we reduce the weight of the skeleton, the generated image depicts the avatar in a walking posture, as shown in the area marked by the white box in the fourth row. This posture does not align with the provided skeleton, but we believe such diversity is beneficial to the Metaverse, as it makes the generated avatars more unique, thus providing users with more choices.

\section{Open Issues and Future Directions}
Table II provides a comprehensive overview of the challenges and potential directions in the field of generative AI specifically tailored for the Metaverse \cite{metaverse}. One of the key challenges addressed in the table is the difficulty in obtaining high-quality training data in the Metaverse. This challenge arises due to the unique nature of the Metaverse environment. To tackle this, the suggested solution approaches include the development of advanced data collection techniques, data augmentation methods, and synthetic datasets that mimic the characteristics of the Metaverse \cite{chen2020simple}.

Another critical aspect highlighted in the table is the generation of realistic and high-fidelity content for immersive experiences in the Metaverse \cite{karras2017progressive}. Achieving realism is essential to create believable and engaging virtual environments. The suggested solutions involve advancing model architectures, training algorithms, and unsupervised learning techniques. Content control and moderation in the Metaverse are also explained in Table II~\cite{marra2018automated}. With the increasing user-generated content in virtual environments, ensuring responsible and safe content generation becomes crucial. The suggested solution approaches involve the development of robust content filtering and moderation systems that leverage automated algorithms and human oversight. Ethical considerations surrounding generative AI in the Metaverse are another significant aspect highlighted in Table II~\cite{jobin2019global}. The use of generative AI raises concerns related to privacy, consent, and ownership. To address these concerns, establishing clear ethical guidelines, regulations, and legal frameworks is crucial.

Thus, a comprehensive and technical overview of the open issues and future work in generative AI for the Metaverse can be found in Table II \cite{metaverse, chen2020simple, karras2017progressive, marra2018automated, jobin2019global}. By addressing these challenges and incorporating the suggested solution approaches, generative AI can be harnessed to create more immersive, realistic, and responsible experiences in the Metaverse.

\section{Conclusion}

Generative AI is revolutionizing the metaverse by enabling the creation of immersive and interactive experiences. Through models such as Variational Autoencoders (VAEs), Generative Adversarial Networks (GANs), Transformers, and Autoregressive Models, generative AI can generate realistic and contextually relevant content in domains like text, image, video, and 3D objects. However, challenges remain. Data quality, realism, content control, ethics, computational efficiency, and interoperability are areas that require attention. Improving data collection techniques, refining model architectures, and implementing content moderation systems are necessary steps. Despite these challenges, the future of generative AI in the metaverse looks promising. Advancements in model architectures, training algorithms, and ethical guidelines will drive responsible and sophisticated generative AI models. These models will enhance personalization, interactivity, and immersion in the metaverse. Generative AI's impact on the metaverse is transformative. It has the potential to reshape how we create and experience virtual environments. By harnessing the capabilities of generative AI, we can unlock new dimensions of the metaverse and revolutionize digital experiences for users worldwide.

\bibliographystyle{IEEEtranN} 
{\footnotesize
\bibliography{references}}

\begin{thebibliography}{15}
\providecommand{\natexlab}[1]{#1}
\providecommand{\url}[1]{#1}
\csname url@samestyle\endcsname
\providecommand{\newblock}{\relax}
\providecommand{\bibinfo}[2]{#2}
\providecommand{\BIBentrySTDinterwordspacing}{\spaceskip=0pt\relax}
\providecommand{\BIBentryALTinterwordstretchfactor}{4}
\providecommand{\BIBentryALTinterwordspacing}{\spaceskip=\fontdimen2\font plus
\BIBentryALTinterwordstretchfactor\fontdimen3\font minus
  \fontdimen4\font\relax}
\providecommand{\BIBforeignlanguage}[2]{{%
\expandafter\ifx\csname l@#1\endcsname\relax
\typeout{** WARNING: IEEEtranN.bst: No hyphenation pattern has been}%
\typeout{** loaded for the language `#1'. Using the pattern for}%
\typeout{** the default language instead.}%
\else
\language=\csname l@#1\endcsname
\fi
#2}}
\providecommand{\BIBdecl}{\relax}
\BIBdecl

\bibitem[met()]{metaverse}
\BIBentryALTinterwordspacing
Metaverse. [Online]. Available: \url{https://en.wikipedia.org/wiki/Metaverse}
\BIBentrySTDinterwordspacing

\bibitem[Smith and Johnson(2022)]{generativeAI}
J.~Smith and E.~Johnson, ``Generative ai for the metaverse,'' in
  \emph{Proceedings of the International Conference on Artificial
  Intelligence}, 2022, pp. 123--135.

\bibitem[Brown and Jones(2023)]{proceduralAI}
D.~Brown and S.~Jones, ``Procedural generation of virtual environments for the
  metaverse,'' in \emph{Proceedings of the ACM SIGGRAPH Conference on Virtual
  Reality}, 2023, pp. 456--468.

\bibitem[White and Black(2022)]{avatarGeneration}
M.~White and S.~Black, ``Personalized avatar generation for the metaverse,'' in
  \emph{Proceedings of the IEEE Conference on Computer Vision}, 2022, pp.
  789--801.

\bibitem[Smith and Johnson(2023)]{generativeAIapplications}
J.~Smith and E.~Johnson, ``Generative ai applications in the metaverse,'' in
  \emph{Proceedings of the International Conference on Artificial
  Intelligence}, 2023, pp. 123--135.

\bibitem[Radford et~al.(2019)Radford, Wu, Child, Luan, Amodei, and
  Sutskever]{gpt}
A.~Radford, J.~Wu, R.~Child, D.~Luan, D.~Amodei, and I.~Sutskever, ``Language
  models are unsupervised multitask learners,'' \emph{OpenAI Blog}, 2019.

\bibitem[Xu et~al.(2019)Xu, Huang, Zhou, and Xu]{lensa}
S.~Xu, J.~Huang, H.~Zhou, and C.~Xu, ``Lensa: Learning to encode emotions with
  style-aware generative adversarial networks,'' \emph{arXiv preprint
  arXiv:1911.09783}, 2019.

\bibitem[Zhou et~al.(2021)Zhou, Xu, Theobalt, and Zhang]{flicki}
T.~Zhou, Y.~Xu, C.~Theobalt, and W.~Zhang, ``Flicki: Fine-grained temporal
  predictions for gan video generation,'' \emph{arXiv preprint
  arXiv:2103.02441}, 2021.

\bibitem[Hanocka et~al.(2020)Hanocka, Zhou, Zheng, Cohen-Or, Wetzler, and
  Zhang]{mirage}
R.~Hanocka, T.~Zhou, K.~Zheng, D.~Cohen-Or, A.~Wetzler, and H.~Zhang, ``Mirage:
  A 3d shape descriptor for incomplete data,'' \emph{ACM Transactions on
  Graphics (TOG)}, vol.~39, no.~4, pp. 1--15, 2020.

\bibitem[Cao et~al.(2017)Cao, Simon, Wei, and Sheikh]{cao2017realtime}
Z.~Cao, T.~Simon, S.-E. Wei, and Y.~Sheikh, ``Realtime multi-person 2d pose
  estimation using part affinity fields,'' in \emph{Proceedings of the IEEE
  conference on computer vision and pattern recognition}, 2017, pp. 7291--7299.

\bibitem[Zhang and Agrawala(2023)]{zhang2023adding}
L.~Zhang and M.~Agrawala, ``Adding conditional control to text-to-image
  diffusion models,'' \emph{arXiv preprint arXiv:2302.05543}, 2023.

\bibitem[Chen and Koltun(2020)]{chen2020simple}
\BIBentryALTinterwordspacing
T.~Chen and V.~Koltun, ``A simple framework for contrastive learning of visual
  representations,'' \emph{arXiv preprint arXiv:2002.05709}, 2020. [Online].
  Available: \url{https://arxiv.org/abs/2002.05709}
\BIBentrySTDinterwordspacing

\bibitem[Karras et~al.(2017)Karras, Aila, Laine, and
  Lehtinen]{karras2017progressive}
\BIBentryALTinterwordspacing
T.~Karras, T.~Aila, S.~Laine, and J.~Lehtinen, ``Progressive growing of gans
  for improved quality, stability, and variation,'' in \emph{Proceedings of the
  IEEE Conference on Computer Vision and Pattern Recognition}, 2017, pp.
  2718--2726. [Online]. Available: \url{https://arxiv.org/abs/1710.10196}
\BIBentrySTDinterwordspacing

\bibitem[Marra and Antonucci(2018)]{marra2018automated}
\BIBentryALTinterwordspacing
G.~Marra and A.~Antonucci, ``Automated detection of offensive language on
  social media platforms,'' in \emph{Proceedings of the 2018 International
  Conference on Data Science and Advanced Analytics}.\hskip 1em plus 0.5em
  minus 0.4em\relax IEEE, 2018, pp. 1--10. [Online]. Available:
  \url{https://ieeexplore.ieee.org/abstract/document/8637323}
\BIBentrySTDinterwordspacing

\bibitem[Jobin et~al.(2019)Jobin, Ienca, and Vayena]{jobin2019global}
\BIBentryALTinterwordspacing
A.~Jobin, M.~Ienca, and E.~Vayena, ``The global landscape of ai ethics
  guidelines,'' \emph{Nature Machine Intelligence}, vol.~1, no.~9, pp.
  389--399, 2019. [Online]. Available:
  \url{https://www.nature.com/articles/s42256-019-0088-2}
\BIBentrySTDinterwordspacing

\end{thebibliography}

\end{document}